# Investigating Warp Size Impact in GPUs

Ahmad Lashgar ¥    Amirali Baniasadi ₮    Ahmad Khonsari ¥

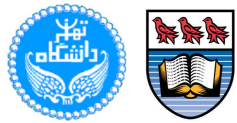

¥ School of Electrical and Computer Engineering
College of Engineering
University of Tehran

₮ Electrical and Computer Engineering Department
University of Victoria

May 2012

This page is left blank intentionally.

# ABSTRACT

There are a number of design decisions that impact a GPU's performance. Among such decisions deciding the right warp size can deeply influence the rest of the design. Small warps reduce the performance penalty associated with branch divergence at the expense of a reduction in memory coalescing. Large warps enhance memory coalescing significantly but also increase branch divergence. This leaves designers with two choices: use a small warps and invest in finding new solutions to enhance coalescing *or* use large warps and address branch divergence employing effective control-flow solutions.

In this work our goal is to investigate the answer to this question. We analyze warp size impact on memory coalescing and branch divergence. We use our findings to study two machines: a GPU using small warps but equipped with excellent memory coalescing (SW+) and a GPU using large warps but employing an MIMD engine immune from control-flow costs (LW+).

Our evaluations show that building coalescing-enhanced small warp GPUs is a better approach compared to pursuing a control-flow enhanced large warp GPU.

# 1. INTRODUCTION

Conventional SIMT accelerators achieve high performance by executing thousands of threads concurrently. In order to simplify GPU design the neighbor threads are bundled in grouped referred to as warps. Employing warp-level granularity simplifies the thread scheduler significantly as it facilitates using coarse-grained schedulable elements. In addition, this approach keeps many threads at the same pace providing an opportunity to exploit common control-flow and memory access patterns. Underlying SIMD units are more efficiently utilized as a result of executing warps built using threads with the same program counter and behavior. In addition, memory accesses of neighbor threads within a warp can be coalesced to reduce the number of off-core requests. Parallel threads overlap the communication overhead associated with some threads using computations required by other threads to maintain high resource utilization.

Previous studies have shown that GPUs are still far behind their potential peak performance as they face two important challenges: branch and memory divergence [6, 11]. Upon branch divergence, threads at one side of a branch stay active while the other side has to become idle. Upon memory divergence, threads hitting in cache have to wait for those who miss. At both divergences, threads suffer from unnecessary waiting periods. This waiting can result in performance loss as leaves the core idle if there are not enough ready threads.

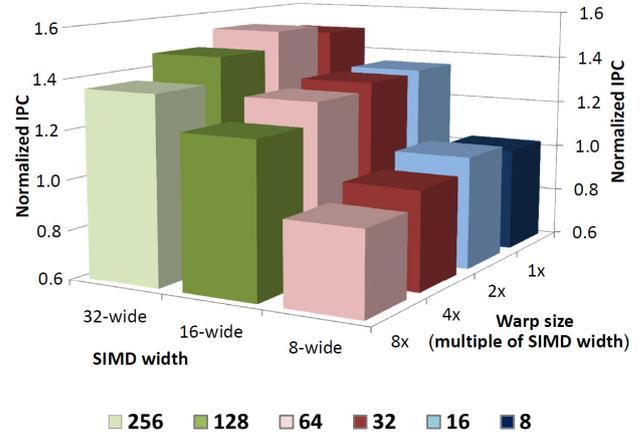

**Figure 1. Warp size impact on performance for different SIMD widths, normalized to 8-wide SIMD and 4x warp size.**

As we show in this work, warp size can impact performance significantly.

*Small warps,* i.e., warps as wide as SIMD width, reduce the likelihood of branch divergence occurrence. Reducing the branch divergence improves SIMD efficiency by increasing the number of active lanes. At the same time, a small size warp reduces memory coalescing, effectively increasing memory stalls. This can lead to redundant memory accesses and increase pressure on the memory subsystem. *Large warps,* on the other hand, exploit potentially existing memory access localities among neighbor threads and coalesce them to a few off-core requests. On the negative side, bigger warp size can increase serialization and the branch divergence impact.

Figure 1 reports average performance for benchmarks used in this study (see methodology for details) for GPUs using different warp sizes and SIMD widths. For any specific SIMD width, configuring the warp size to 1-2X larger than SIMD width provides best average performance. Widening the warp size beyond 2X degrades performance. In the remainder of this paper, we use an 8-wide SIMD configuration.

In this paper we analyze how warp size impacts performance in GPUs. We start with studying GPUs using different warp sizes ranging from 8 to 64. We use our analysis to investigate the effectiveness of two possible approaches to enhance GPUs. The first approach relies on enhancing memory coalescing in GPUs using large warps. Once memory coalescing is enhanced, this approach uses effective control-flow solutions to address the resulting increase in branch divergence. The second approach aims at minimizing serialization in GPUs using small warps. Since small warps affect coalescing negatively, this approach requires taking extra steps to address memory stalls. We may expect the two approaches to be equally effective as they address GPU's performance degrading issues, memory and branch divergence, simultaneously. However, our experimental results show that often one outperforms the other.

In this work we evaluate both approaches and estimate the performance return of both solutions. We show that starting with a small warp size, and then using dynamic memory divergence solutions is a better choice.

In summary we make the following contributions:

= We study the impact of warp size on different GPU aspects including memory stalls, idle cycles and performance.
= We use our findings to identify an effective approach to enhancing GPU performance. We show that the combination of a static and simple approach to deal with branch divergence (using small warps) and dynamic memory stall reductions solutions is an effective approach.
= We also investigate the alternative and show that using a static solution to enhance coalescing (i.e., using large warps) combined with an ideal dynamic control-flow solutions falls behind the first approach due to frequent synchronization of a large number of threads.

The rest of the paper is organized as follows. In Section 2 we study background. In Section 3 we review warp size impact. In Section 4 we present our machine models. In Section 5 we discuss methodology. Section 6 reports results. In Section 7 we discuss our findings in more detail. In Section 8 we review related work. Finally, Section 9 offers concluding remarks.

## 2. BACKGROUND

In this study we focus on SIMT accelerators similar to NVIDIA Tesla [12]. Stream Multiprocessors (SMs) are processing cores sending memory requests to memory controllers through on-chip crossbar network. We augment Tesla with private L1 caches for each SM.

Each SM keeps context for 1024 threads. SM has one thread scheduler which groups and issues warps on one SIMD group. Threads within a warp have one common program counter. Control-flow divergence among threads is managed using re-convergence stack [6]. Diverged threads are executed serially until re-converging at the immediate post-dominator of branch.

Instructions from different warps are issued back-to-back into the deep 24-stage, 8-wide SIMD pipeline. If the warp pool has no ready warp, the pipeline front-end stays idle leading to underutilization. Under such circumstances, all the warps are issued into the pipeline. However, there are ready threads that are inactive/waiting due to branch/memory divergence [13].

Current GPUs coalesce the global memory accesses of neighbor threads. We model a coalescing behavior similar to compute compatibility 2.0 [15]. Requests from neighbor threads accessing the same stride are coalesced into one request. Neighbor threads are aggregated over the entire warp. Consequently, memory accesses of a warp are coalesced into one or many stride accesses. Each stride is 64 bytes. Memory transaction granularity is the same as cache block size which is one stride.

## 3. WARP SIZE IMPACT

In this section we report how warp size impacts, the number of idle cycles, memory access coalescing, and performance. We do not report SIMD efficiency since our observations show that warp size has insignificant (less than 1%) impact on activity factor ([10]). See Section 5 for methodology.

**Memory access coalescing.** Memory accesses made by threads within a warp are coalesced into fewer memory transactions to reduce bandwidth demand. We measure memory access coalescing using the following equation:

$$Coalescing\ rate = \frac{Total\ off chip\ requests}{Total\ memory\ insn.} \quad (1)$$

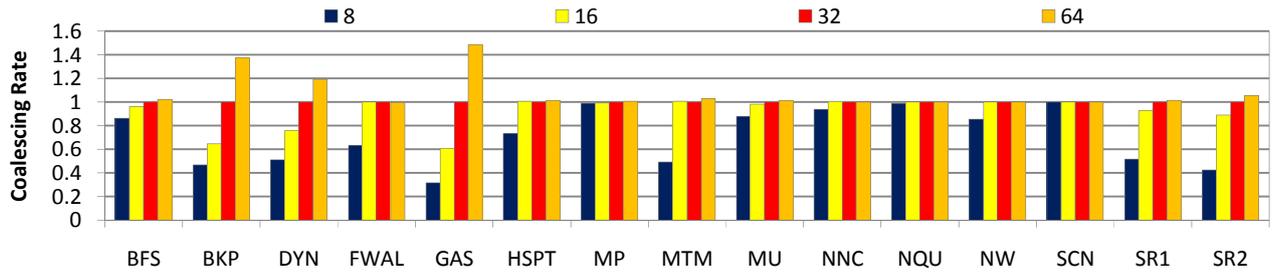

Figure 2. Coalescing rate for GPUs using different warp sizes, normalized to a GPU using 32 threads per warp.

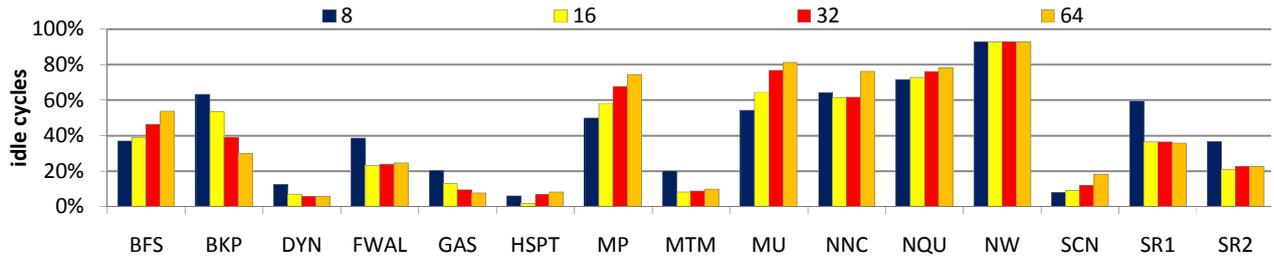

Figure 3. Idle cycle share for GPUs using different warp sizes.

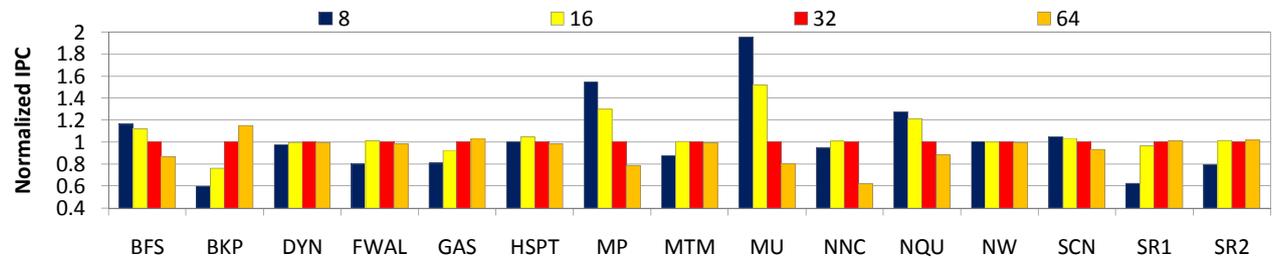

Figure 4. Performance for GPUs using different warp sizes.

We use this definition (equation 1) to estimate coalescing in different machine models studied in this paper. Figure 2 compares coalescing rates for different warp sizes. As shown in the figure, increasing the warp size improves the coalescing rate in all the benchmarks. An increase in warp size can increase the likeliness of memory accesses made to the same cache block residing in the same warp. This increase starts to diminish for warp sizes beyond 32 threads for most benchmarks as coalescing width (16 words of 32-bit) becomes saturated. Accordingly, enlarging the warp beyond a specific size, returns little coalescing gain.

**Idle cycles.** Figure 3 reports idle cycle frequency for GPUs using different warp sizes. Idle cycles are cycles when the scheduler finds no ready warps in the pool. Core idle cycles are partially the result of branch/memory divergences which inactivate otherwise ready threads [13]. Small warps compensate branch/memory divergence by hiding idle cycles (e.g. BFS). On the other hand, for some benchmarks (e.g., BKP), small warps lose many coalescable memory accesses increasing memory pressure. This pressure increases average core idle durations compared to larger warps (e.g. BKP).

**Performance.** An increase in warp size can have opposite impacts on performance. Performance can improve if an increase in memory access coalescing compensates synchronization overhead imposed by large warps. Performance can suffer if the synchronization overhead associated with large warps outweighs coalescing memory access gains. Figure 4 reports performance for GPUs using different warp sizes. As reported, in most workloads, warp size has significant impact on performance. Performance improves in BKP, GAS, SR1 and SR2 with warp size. Performance is lost in BFS, MP, MU, NQU and SCN as warp size increases. Other workloads perform best under average warp sizes (16 or 32 threads).

We conclude from this section that warp size can impact performance in different ways. We use our findings and explore two approaches to enhance performance further in GPUs. We describe our approaches in the next section.

## 4. MACHINE MODELS

In this section we introduce two machine models. Our first model is a coalescing-enhanced small warp machine, referred to as SW+. SW+ uses small warps but comes with ideal coalescing. Intuitively we study SW+ to measure the performance potential in building small warp machines. Our second model represents a control-flow-enhanced large warp machine, referred to as LW+. We use LW+ to estimate the performance improvement possible for a processor using a large warp size but equipped with an ideal control-flow solution.

**Table 1. Baseline configurations for GPGPU-sim.**

| NoC | |
|---|---|
| Total Number of SMs | 16 |
| Number of Memory Ctrls | 6 |
| Number of SM Sharing an Network Interface | 2 |
| Network Interface Buffer size | 8 |
| **SM** | |
| Warp Size | 32 Threads |
| Number of thread per SM | 1024 |
| Maximum allowed CTA per SM | 8 |
| Number of register per SM | 16K 32-bit |
| SM SIMD width | 8 |
| Shared Memory size | 16KB |
| L1 Data cache | 48KB : 8-way : LRU : 64BytePerBlock |
| L1 Texture cache | 16KB : 2-way : LRU : 64BytePerBlock |
| L1 Constant cache | 16KB : 2-way : LRU : 64BytePerBlock |
| **Clocking** | |
| Core clock | 1300 MHz |
| Interconnect clock | 650 MHz |
| DRAM memory clock | 800 MHz |
| **Memory** | |
| Number of Banks Per Memory Ctrls | 8 |
| DRAM Scheduling Policy | FCFS |
| GDDR3 Memory Timing | tRRD=12, tRCD=21, tRAS=13, tRP=34, tRC=9, tCL=10 |

**Table 2. Benchmarks Characteristics.**

| Name | Grid Size | Block Size | #Insn |
|---|---|---|---|
| **BFS**: BFS Graph[3] | 16x(8,1,1) | 16x(512,1,1) | 1.4M |
| **BKP**: Back Propagation[3] | 2x(1,64,1) | 2x(16,16,1) | 2.9M |
| **DYN**: Dyn_Proc[3] | 13x(35,1,1) | 13x(256,1,1) | 64M |
| **FWAL**: Fast Walsh Transform[6] | 6x(32,1,1) 3x(16,1,1) (128,1,1) | 7x(256,1,1) 3x(512,1,1) | 11.1M |
| **GAS**: Gaussian Elimination[3] | 48x(3,3,1) | 48x(16,16,1) | 8.8M |
| **HSPT**: Hotspot[3] | (43,43,1) | (16,16,1) | 76.2M |
| **MP**: MUMmer-GPU++[8] | (1,1,1) | (256,1,1) | 0.3M |
| **MTM**: Matrix Multiply[14] | (5,8,1) | (16,16,1) | 2.4M |
| **MU**: MUMmer-GPU[1] | (1,1,1) | (100,1,1) | 0.15M |
| **NNC**: Nearest Neighbor on cuda[2] | 4x(938,1,1) | 4x(16,1,1) | 5.9M |
| **NQU**: N-Queen [1] | (256,1,1) | (96,1,1) | 1.2M |
| **NW**: Needleman-Wunsch [3] | 2x(1,1,1) … 2x(31,1,1) (32,1,1) | 63x(16,16,1) | 12.9M |
| **SC**: Scan[14] | (64,1,1) | (256,1,1) | 3.6M |
| **SR1**: Speckle Reducing Anisotropic Diffusion [3] (large dataset) | 3x(8,8,1) | 3x(16,16,1) | 9.1M |
| **SR2**: Speckle Reducing Anisotropic Diffusion [3] (small dataset) | 4x(4,4,1) | 4x(16,16,1) | 2.4M |

### 4.1 SW+

This machine exploits small warps (as wide as SIMD width).

As described before, small warps lose some coalescing opportunities leading to redundant memory accesses. SW+ is enhanced to address the performance penalty associated with uncoalesced accesses. SW+ is equipped with ideal coalescing hardware, which coalesces the memory accesses of all threads. Ideal coalescing hardware keeps track of outstanding memory requests (of all threads) and merges read accesses with outstanding accesses whenever possible. This merging captures most coalescing opportunities occurring for large warps, compensating the penalty paid by small warps effectively.

Our baseline architecture coalesces accesses within one warp. SW+ extends coalescing beyond one warp.

The motivation behind investigating SW+ is to study if investing in a small warp size machine to enhance memory coalescing can lead to high performance returns.

### 4.2 LW+

We investigate LW+ to evaluate if investing in a large warp size machine to enhance branch divergence is the right approach. LW+ groups threads in large warps (8x larger than the SIMD width). Exploiting large warps facilitates efficient usage of memory bandwidth by coalescing memory accesses.

Large warps exacerbate idle periods imposed by branch divergence. LW+ addresses this issue as both sides of divergence are split and actively remain in the warp pool in this machine. This splitting does not return considerable performance gain since threads may never re-converge again leading to SIMD underutilization [7]. Therefore, we further enhance this machine by replacing the SIMD lanes with MIMD cores. Splitting upon divergence and using MIMD cores solves both problems.

Previous studies have suggested solutions to reduce the impact of branch divergence [6, 7, 14, 2]. DWS adaptively splits the warp upon branch/memory divergence [13]. DWF, TBC, LWM, SBI and SWI propose solutions to capture a considerable amount of MIMD performance by SIMD. Exploiting DWS on top of TBC or LWM can be viewed as a practical approach in building LW+-like processors.

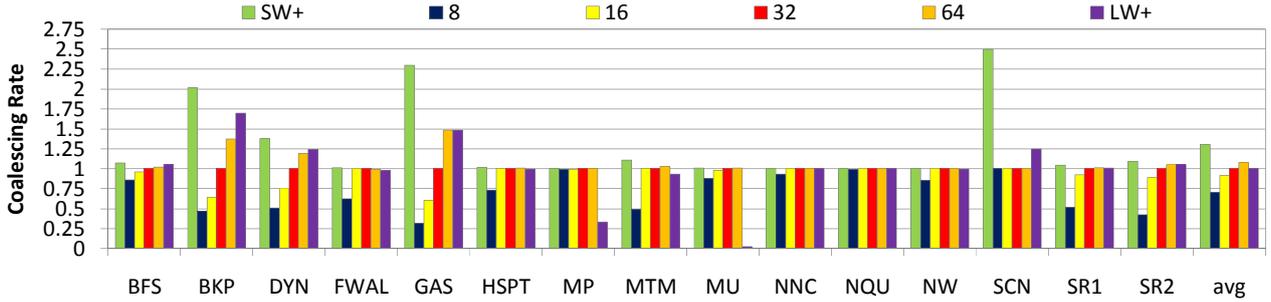

Figure 5. Coalescing rate for SW+, LW+ and processors using different warp sizes.

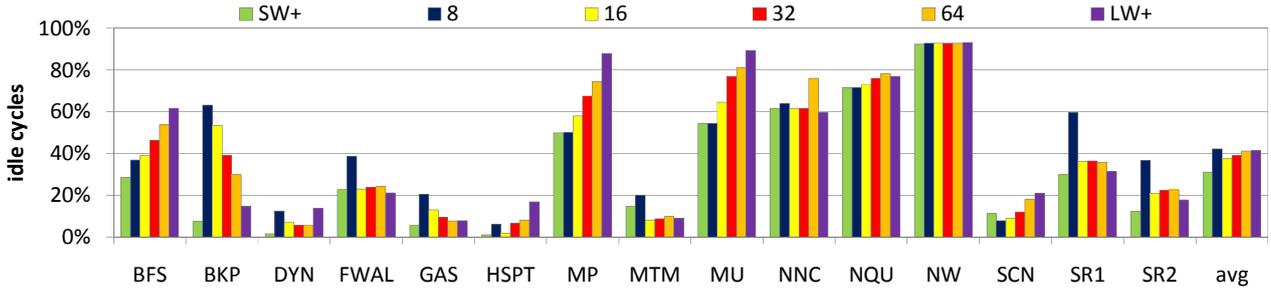

Figure 6. Idle cycle share for SW+, LW+ and processors using different warp sizes.

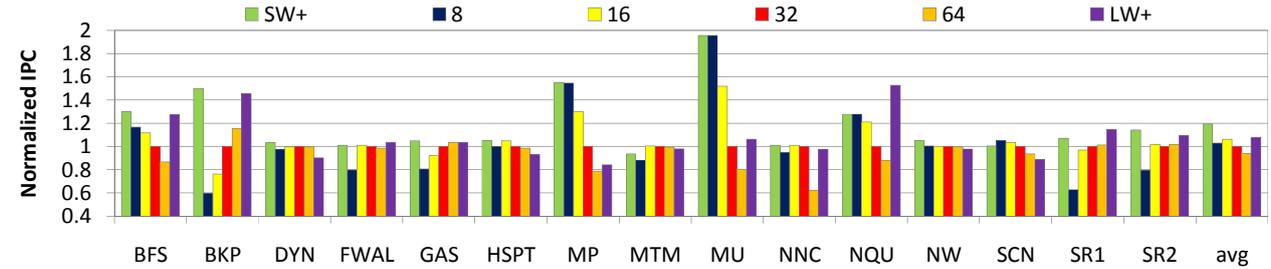

Figure 7. Performance for SW+ and LW+, processors using different warp sizes.

## 5. METHODOLOGY

We modified GPGPU-sim [1] (version 2.1.1b) to model large warps and memory coalescing similar to compute compatibility 2.0 devices [16]. We used the configurations shown in Table 1 to model the baseline microarchitecture described in Section 2. 16 SMs provide peak throughput of 332.8 Gflops. Six 64-bit wide memory partitions provide memory bandwidth of 76.8 Gbytes/s at dual-data rate.

We used a cache block size of 64 bytes, which is equal to memory transaction chunks. Our evaluations show increasing cache block size (and accordingly transaction chunk) to 128 bytes, downgrades the overall performance.

We used benchmarks from GPGPU-sim [1], Rodinia [3] and CUDA SDK 2.3 [15]. We also included MUMmerGPU++ [8] third-party sequence alignment program. We use benchmarks exhibiting different behaviors: memory-intensiveness, compute-intensiveness, high and low branch divergence occurrence and with both large and small number of concurrent thread-block. Table 2 shows our benchmarks and their characteristics.

## 6. RESUTLS

In this section, we evaluate SW+, LW+ and processors using different warps sizes. In Section 6.1 we present memory access coalescing. SIMD efficiency is reported in 6.2. Finally in Section 6.3 we report performance.

### 6.1 Memory access coalescing

Figure 5 reports coalescing rate. As reported, SW+ provides the best coalescing rate. SW+ coalesces memory accesses among all threads of an SM to achieve this. Widening the coalescing to merge accesses from all threads can improve coalescing rate by 21% and 30% compared to coalescing width of 32 threads and 64 threads, respectively.

LW+ is outperformed by a machine using 64 threads per warp. This is due to the fact that LW+'s MIMD execution does not keep threads at the same pace to coalesce their accesses. In some cases (e.g., MP and MU) splitting the warp upon divergence prevents merging memory requests. Under such circumstances, redundant memory accesses lead to poor coalescing rate. As we show later, this does not translate to performance loss since the memory subsystem is not under-pressure in these workloads (MU and MP).

## 6.2 Idle cycles

As discussed in Section 2, small warps reduce idle cycles by reducing unnecessary waiting due to branch/memory divergence. This idle cycle saving is lost partially since small warps lose memory access coalescing, pressuring the memory subsystem. SW+ addresses this drawback by exploiting ideal coalescing. As shown in Figure 6, SW+ shows lowest idle cycle share in most workloads. On average, using short warps combined with ideal coalescing (SW+), reduces idle cycles by 36%, 21% and 26% compared to processors using 8, 16 and 32 threads per warp, respectively.

Our analysis shows synchronizing a large number of threads at every instruction increases the number of idle cycles in LW+ significantly.

## 6.3 Performance

Figure 7 reports performance for SW+, LW+ and processors using different warp sizes. SW+ outperforms all alternatives in most benchmarks. On average, SW+ outperforms LW+, and machines using 8, 16 and 32 threads per warp by 11%, 16%, 12% and 19%, respectively.

LW+ synchronizes all threads of the warp at every instruction. Even MIMD cores cannot compensate this synchronization overhead. Therefore a big part of MIMD's gain is lost due to unnecessary waitings. On average, LW+ outperforms processors using 8, 16, 32 and 64 threads per warp by 5%, 1%, 7% and 15%, respectively.

## 7. DISCUSSION

In this section we comment on some practical implications and analyze our results further.

**Insensitive workloads.** Warp size affects performance in SIMT cores only for workloads suffering from branch/memory divergence or showing potential benefits from memory access coalescing. Therefore, benchmarks lacking either of these characteristics (e.g. FWAL and DYN) are insensitive to warp size.

**Enhancing short warps.** Among all configurations, a GPU using 8 threads per warp performs worst for many benchmarks (e.g., BKP) as it suffers from very low memory coalescing. SW+'s investment in addressing this issue comes with considerable (up to 95%) returns. However, this machine performs well for computation-bounded benchmarks (e.g. BFS, MP, MU and NQU), which suffer from branch divergence significantly.

**Enhancing large warps.** A closer look at the processor using 64 threads per warp shows that it performs well for a few benchmarks (e.g. BKP, GAS and SR1 and SR2), but falls behind for BFS, MU, MP, NNC, NQU and SC which are prone to branch divergence. Enhancing this processor with an effective control-flow solution, however, shows very high (up to a maximum of 73% in NQU) performance returns.

**Ideal coalescing and write accesses.** SW+'s coalescing rate is far higher than other machines due to ideal coalescing hardware. However, ideal coalescing can only capture the read accesses and does not compensate un-coalesced accesses. Therefore, SW+ may suffer from un-coalesced write accesses. We found this rare as it can be seen only in the MTM benchmark. The coalescing rate of SW+ in MTM is higher than other machines since it merges many read accesses among warps. However, un-coalesced write accesses downgrades the overall performance in SW+.

**Practical issues with small warps.** Pipeline front-end includes the warp scheduler, fetch engine, decode instruction and register read stages. Using fewer threads per warp affects pipeline front-end as it requires a faster clock rate to deliver the needed workload during the same time period. An increase in the clock rate can increase power dissipation in the front-end and impose bandwidth limitation issues on the fetch stage. Moreover, using short warps can impose extra area overhead as the warp scheduler has to select from a larger number of warps. In this study we focus on how warp size impacts performance, leaving the area and power evaluations to future works.

**Register file.** Warp size affects register file design and allocation. GPUs allocate all warp registers in a single row [5]. Such an allocation allows the read stage to read one operand for all threads of a warp by accessing a single register file row. For different warp sizes, the number of registers in a row (row size) varies according to the warp size to preserve accessibility. Row size should be wider for large warps to read the operands of all threads in a single row access and narrower for small warps to prevent unnecessary reading.

## 8. RELATED WORKS

Kerr et al. [10] introduced several metrics for characterizing GPGPU workloads. Bakhoda et al. [1] evaluated the performance of SIMT accelerators for various configurations including interconnection networks, cache size and DRAM memory controller scheduling. Lashgar and Baniasadi [11] evaluated the performance gap between realistic SIMT cores and semi-ideal GPUs to identify appropriate investment points.

Dasika et al. [4] studied SIMD efficiency according to the SIMD width. Their study shows the frequent occurrence of divergence in the scientific workloads makes wide SIMD organizations inefficient in terms of performance/watt. 32-

wide SIMD is found to be the most efficient design for the studied scientific computing workloads.

Jia et al. [9] introduced a regression model relating the GPU performance to microarchitecture parameters such as SIMD width, thread block per core and shared memory size. Their study did not cover warp size but concluded that SIMD width is the most influential parameter among the studied parameter.

## 9. CONCLUSION

Filling the performance gap between current GPUs and their potential requires addressing both memory and branch divergence.

Finding the right configuration of a GPU is perhaps the most important decision in achieving high performance. Such static decisions, however, influence the dynamic solutions a system requires to deal with runtime challenges. Choosing the right warp size is one example. Approaching memory coalescing with a static solution (using a large size warp), leaves us with the challenge of finding effective dynamic control-flow solutions. An alternative approach is to deal with control-flow first by using small warps and then investigating dynamic solutions to address memory coalescing.

We study the performance potential for both approaches and conclude that the latter approach comes with better performance returns for benchmarks and configurations used in this work.